\documentstyle[titlepage,fleqn,11pt]{article}
\begin{document}
\begin{titlepage}
\title{Electromagnetic Modes of Maxwell Fisheye Lens}
\author{Haret Rosu\\
\\
Instituto de F\'{\i}sica, Universidad de Guanajuato,\\
Apdo Postal E-143, 37150 Le\'on, Gto, M\'exico\\
\\
Marco Reyes\\
\\
Departamento de F\'{\i}sica, CINVESTAV, Apdo Postal 14-740,\\
07000 M\'exico D.F., M\'exico\\
 }

\date{}
{\baselineskip=20pt

\begin{abstract}

We provide an analysis of the radial structure of TE and TM modes of the
Maxwell fisheye lens, by means of Maxwell equations as applied to the
fisheye case. Choosing a lens of size R = 1 cm, we plot some of
the modes in the infrared range.

\vskip 2cm

Pacs Numbers: 03.50.De, 42.79.Ry

\vskip 1cm

internet: rosu@ifug.ugto.mx

\end{abstract}

  }

 \maketitle
 \end{titlepage}
\baselineskip=25pt
\parskip=0pt

 Maxwell fisheye lenses $^{ \cite{tb}}$
 are spherical inhomogenous spheres which
belong to the class of complete, perfectly focussing optical
systems, that is, within geometrical optics, (i) all rays are circles
lying in planes containing the centre of the lens (ii) every point $P$
has a conjugate $P^{'}$ (iii) the imaging is an inversion,
$OP\times OP^{'}= R^{2}$ (there are no spherical aberrations).
Due to such good geometrical qualities, it is no wonder to be used
in animal vision. At the present time people are trying to reproduce
fisheye lenses by the so-called GRIN (graded index) technology
$^{\cite{gr}}$.
The fisheye first appeared in {\em Physics} in the past century as a
problem of the Irish Academy looking for the refractive index of a
medium that could conceivably form images in the least depth. It was
solved by Maxwell in 1854. The fisheye lens is one of
the most
symmetric systems one could find in Nature, and has been called "the
hydrogen atom of optics" $^{\cite{0}}$. Most studies considered only the
geometrical optics approximation (eikonal theory). The similarities
to the Kepler/Coulomb problem have been pointed out a
long time ago $^{\cite{kep}}$.
These lenses have also interesting connections
with phenomenological electrodynamics in curved space.
Moreover, one can draw an analogy between the optical properties of
standard Maxwell lenses and those of the synchronous gauge
$k=1$ of de Sitter spacetime (also called de Sitter-Lanczos model) in
its expanding stage $^{\cite{001}}$. Besides, there are many interesting
 open problems concerning the propagation of signals in inhomogeneous
 media that were recently touched upon in the literature $^{\cite{sig}}$.
Due to their aberration-free imaging over a wide range of angles,
inhomogeneous spherical lenses are important for antennas such as required
in radar technology.

In 1958, Tai $^{\cite{t}}$ discussed the Maxwell equations in the
fisheye case. At the end of his paper, Tai complained about the missing
of computers allowing detailed
numerical studies of his theoretical treatment. After many years, our work
is just a step
in this direction, complemented with some more comments.
Tai parametrized the spherical symmetric index of refraction in
the standard/Maxwell form
$$\kappa(r)=\frac{2}{1+(r/R)^{2}}    \eqno(1)$$
It is well-known that the electromagnetic field inside a spherically
symmetric but radially inhomogeneous medium can always be expressed as
a sum of TE and TM modes $^{\cite{t2}}$ by an
extension of the vector wavefunction method $^{\cite{st}}$.

With the parametrization Eq.(1), one can obtain the following
two equations, expressing the radial modes of the lens
 $$
 \frac{d^{2}E_{n}}{dr^{2}}+[k^{2}\kappa ^{2}(r)-\frac{n(n+1)}{r^{2}}]E_{n}
  = 0  \;\; \eqno(2) $$
  $$
 \frac{d^{2}M_{n}}{dr^{2}}-\frac{1}{\kappa ^{2}(r)}\frac{d\kappa ^{2}(r)}{dr}
 \frac{dM_{n}}{dr}+[k^{2}\kappa ^{2}(r)-\frac{n(n+1)}{r^{2}}]M_{n} =0
 \eqno(3)$$
 where $k$ is the wave number and $n$ is an integer. The two functions
 $E_{n}$ and $M_{n}$ correspond , respectively, to the transverse
  electric and transverse magnetic fisheye modes. The above equations
 are turned into hypergeometric differential equations for
  two functions denoted $U$ and $V$ as follows
  $$\xi(\xi-1)U^{''}+[(2\mu +\beta)\xi-\beta]U^{'}+\alpha U=0
   \eqno(4)$$
   and
   $$\xi(\xi -1)V^{''}+[(2\mu +\beta)\xi -\beta]V^{'} +(\alpha -1/2)V=0
     \eqno(5)$$
where
     $$\alpha=\beta\mu+(kR)^{2};\;\; \beta=n+3/2;\;\; \mu=1/2(1+
     \sqrt{1+(2kR)^{2}})   \eqno(6)$$
with derivatives taken with respect to $\xi=-(r/R)^{2}$.

The pair of functions $(U,V)$ is related to the other pair
($E_{n},M_{n}$) in the following way:
$$E_{n}=\xi ^{(n+1)/2} (\xi -1)^{\mu}U    \eqno(7)$$
$$M_{n}=\xi^{(n+1)/2}(\xi -1)^{\mu -1}V   \eqno(8)$$

The functions $U$ and $V$ are
hypergeometric functions of 3 parameters F(a,b,c;$\xi$) with a,b,c
identified in terms of $\alpha$, $\beta$, and $\mu$. The needed
hypergeometric functions should be the ones regular at the origin
$\xi=0$. Thus we discarded the second independent solution
having a $\xi ^{-(n+1)/2}$ singularity at the origin $^{\cite{arf}}$.

After the corresponding identification of the parameters, the most
convenient form of the $U$ and $V$ functions is either
$$U=F_{U}(\mu,\mu +n+1/2, n+3/2;\xi)   \eqno(9)$$
or a second $U$ function with the first two parameters interchanged,
and
$$V=F_{V}(\mu+(2n+1+2\sqrt{(n+1/2)^{2}+2})/4,
\mu+(2n+1-2\sqrt{(n+1/2)^{2}+2})/4,n+3/2;\xi)    \eqno(10)$$
or a second $V$ function again with the first two parameters
interchanged.

Equations (9) and (10) are the analytical expressions for the modal
functions (up to two power factors) of the Maxwell fisheye in the
standard parametrization of the refractive index.
The two functions $U$, with the first two parameters interchanged,
express only a single function as one can easily
see from the power expansion of hypergeometric functions or from
integral representations $^{\cite{grr}}$.
The same happens for the pair of $V$ functions, and therefore
we work only with the forms given by Eq. (9) and Eq. (10).

The analytical functions we want to plot are:
$$E_{n}=(-\xi) ^{(n+1)/2}(1-\xi)^{\mu} F_{U}   \eqno(11)$$
and
$$M_{n} =(-\xi) ^{(n+1)/2} (1- \xi)^{\mu -1} F_{V} \eqno(12) $$
which are identical to Eq. (7) and Eq. (8), respectively.

We choose to plot some of the modes in the IR wavelength range. The
results are shown in Figs. 1, 2, and 3, where both the one -dimensional
radial plots and the three-dimensional pictures are displayed.

We finally recall that the geometrical approximation requires the
following condition to be satisfied
$$\frac{\nabla \kappa(r)/ \kappa(r)}{k\kappa(r)}\ll 1     \eqno(13)$$
where $k$ is the free-space wave number. The spatial dispersion effects in
the fisheye lens result in the following relation:

$$\frac{\nabla n(r)}{kn^{2}(r)}=|-\frac{r}{R}\cdot \frac{1}{kR}| \ll 1
\eqno(14)$$
when the parametrization given by Eq. (1) is used. This condition is well
satisfied for the wavelength range considered by us.
The difference between
the ray theory and Maxwell equations in the case of the Maxwell lens is
merely that while the first predicts the propagation of light along circles,
the
second gives a distribution of the electromagnetic field over the circles.

\section*{ Acknowledgments}

This work was supported partially by CONACyT Grant No. F246-E9207
to the University of Guanajuato, and by a CONACyT Graduate Fellowship.


\section*{Figure Captions}

\begin{itemize}

\item {\bf Fig. 1} : a) and b), E1 and M2 modes for $kR=10$, (IR range)
                     respectivelly;
                     c) and d), the same modes for $kR=100$, (middle IR range)

\item {\bf Fig. 2} : 3-dimensional plots for the same cases.

\item {\bf Fig. 3} : E3 mode for $kR=1000$, (near IR range)

\end{itemize}

\end{document}